\documentclass{fic-l}

\newtheorem{theorem}{Theorem}[section]

\theoremstyle{definition}

\theoremstyle{remark}

\numberwithin{equation}{section}

\usepackage{graphicx,amssymb,epsf}

\newcommand{\be}{\begin{equation}}
\newcommand{\ee}{\end{equation}}
\newcommand{\bea}{\begin{eqnarray}}
\newcommand{\eea}{\end{eqnarray}}
\newcommand{\beas}{\begin{eqnarray*}}
\newcommand{\eeas}{\end{eqnarray*}}

\def\One{\mathbb{I}}

\def\v{\;\raisebox{-0mm}{\epsfysize=4mm\epsfbox{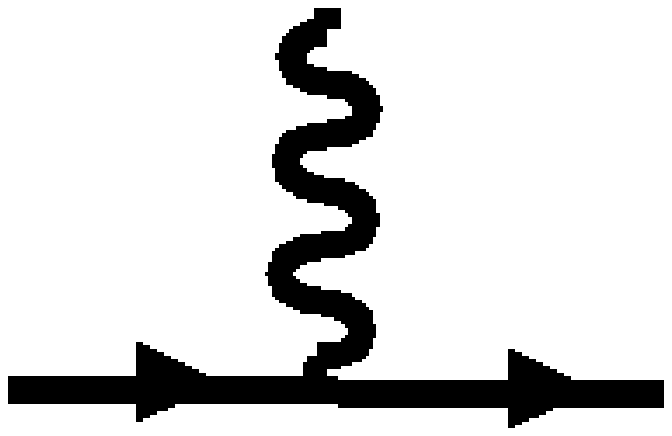}}\;}
\def\f{\;\raisebox{-0mm}{\epsfysize=2mm\epsfbox{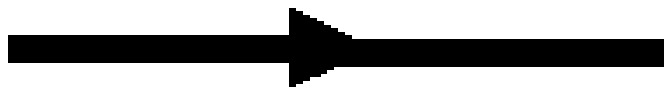}}\;}
\def\g{\;\raisebox{-0mm}{\epsfysize=2mm\epsfbox{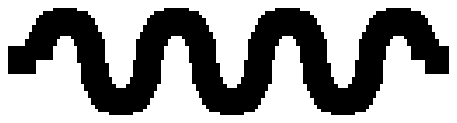}}\;}

\def\qeddse{\;\raisebox{-16mm}{\epsfysize=36mm\epsfbox{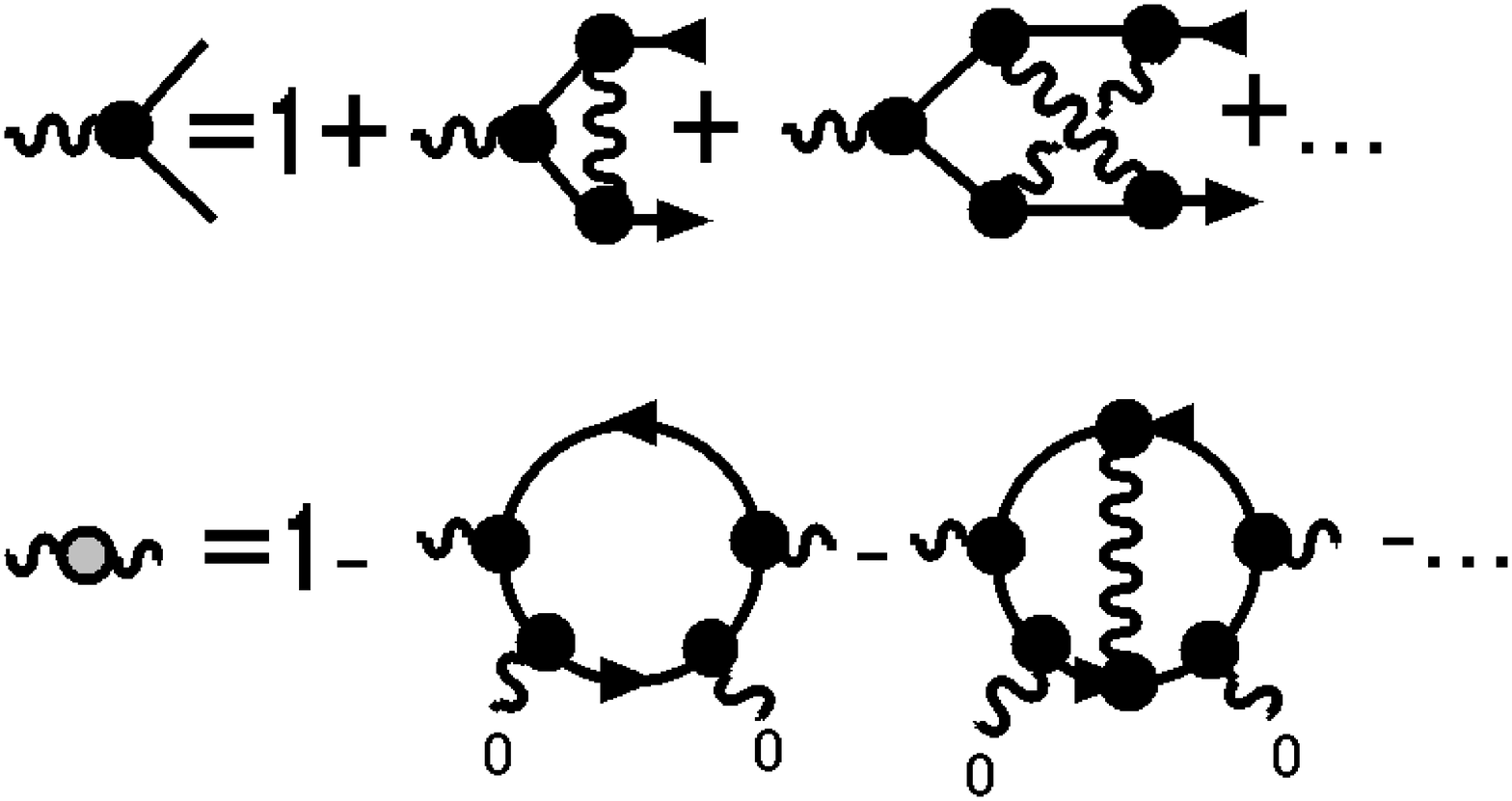}}\;}

\begin{document}
\title{Dyson--Schwinger Equations: From Hopf algebras to Number Theory
}
\author{Dirk Kreimer}
\address{IH\'ES\\ Le Bois Marie, 91440 Bures sur Yvette, France\\ kreimer@ihes.fr}
\thanks{supported by CNRS and by grant NSF-DMS0603781 at Center for Mathematical Physics,
Boston University. email:  kreimer@ihes.fr}



\begin{abstract}
We consider the structure of renormalizable quantum field theories
from the viewpoint of their underlying Hopf algebra structure. We
review how to use this Hopf algebra and the ensuing Hochschild
cohomology to derive non-perturbative results for the short-distance
singular sector of a renormalizable quantum field theory.  We focus
on the short-distance behaviour and thus discuss renormalized Green
functions $G_R(\alpha,L)$ which depend on a single scale $L=\ln
q^2/\mu^2$.
\end{abstract}

\maketitle

\section{Introduction}
The crucial notion of locality, the structure of Dyson--Schwinger
equations and the appearance of mixed motives in the evaluation of
Feynman graphs are intimately related. We want to exhibit how these
notions come together in quantum field theory. We emphasize the role
of Dyson--Schwinger equations in this interplay.

Renormalization theory is a time-tested subject put to daily use in
many branches of physics. We have seen many of its facets
illuminated here at the Fields Institute. In this paper, we focus on
its applications in quantum field theory, where a standard
perturbative approach is provided through an expansion in Feynman
diagrams. In perturbation theory it is mainly a combinatorial
problem: determine the needed correction to parameters in the
Lagrangian such that the computation allows for finite results in
the desired order of perturbation. Whilst the resulting
combinatorics of the Bogoliubov recursion, solved by suitable forest
formulas, has been known for a long time, the subject regained
interest on the conceptual side with the discovery of an underlying
Hopf algebra structure behind these recursions.

Non-perturbatively, one faces the equations of motion which the full
Green functions have to fulfill. These Dyson--Schwinger equations
(DSEs) reflect the self-similarity of amplitudes in quantum field
theory, upon studying their skeleton expansion: the computation of
propagation or interaction of amplitudes proceeds by taking into
account that the same propagation or interaction can happen in
internal processes. Hence the notion of internal process  demands
objects which possess internal structure. These turn out to be the
celebrated skeleton graphs of a theory. They are crucial in
understanding at the same time the algebraic as well as the
number-theoretic properties of a field theory.

The main goal of this review is to emphasize one crucial point: How
the algebraic structure of the perturbative expansion can lead to
non-perturbative solutions, by emphasizing the unifying role of
algebraic structures in understanding local renormalization as well
as the above self-similar structure of Green functions.

We will also point out along the way how internal symmetries are
reflected in this set-up. This is a very recent insight
\cite{anatomy}, which we can only mention in passing. It goes a long
way in establishing a quantum field theory which stands its own
ground: instead of deriving quantum gauge invariance from the
differential geometry of a classical gauge field theory, we obtain
it as a consequence of the algebraic structures of a local quantum
theory, and the classical Lagrangian is a derived quantity. It is
indeed obtained from the residues of Feynman graphs which are
primitive elements in the Hopf algebra of renormalization, an old
result in field theory \cite{instanton}.

Finally, in the course of our discussion we will give an idea where
one can see signs of universality from this set-up. It is at the
time of writing an open research problem to connect the rather
systematic account to DSEs exhibited in this talk to other areas
where renormalization theory and universality are likewise
fundamental notions, as these proceedings hopefully demonstrate.

We first have to review the algebraic structure of quantum field
theory, and with it the Hopf algebra structure of a perturbative
expansion. From a strictly perturbative viewpoint, this is
summarized in this volume in the paper by Ebrahimi-Fard and Guo
\cite{GF}, with emphasis given to Rota--Baxter algebras. Our own
summary below follows \cite{Enc}. We hence will be short in our
discussion of the perturbative viewpoint, and aim at a qualitative
discussion of the above points, exemplified in the much stressed
example of the propagator in massless Yukawa theory \cite{BrK,KrY}.
Hochschild cohomology then leads the way from perturbative physics
to non-perturbative results. We summarize here results detailed in
previous works on the structure of DSEs
\cite{anatomy,BrK,KrY,BergbKr,houches,linear} and will end in
discussing shortly the number-theory of skeleton graphs, which found
a mathematical interpretation as periods of mixed motives recently
\cite{BEK}. We will be content in exhibiting how those primitive
graphs mentioned above provide the connection to algebraic geometry
and motives.

\subsection{A splitting of amplitudes}
We assume we work in a renormalizable quantum field theory which
provides a finite set ${\mathcal R}\subset {\mathcal A}$  of
amplitudes which need renormalization. Here, ${\mathcal A}$ is the
set of all amplitudes in a given theory. We hence work with a theory
which, from a Lagrangian perspective, provides a finite number of
parameters which need renormalization. Note that our notion of an
amplitude is such that each monomial in said Lagrangian corresponds
to an amplitude. In comparison with the standard terminology, we
distinguish the various form-factors provided by a given Green
function and regard each corresponding structure function as a Green
function. For example the vertex function for the photon decay into
an electron-positron pair in quantum electrodynamics (QED) has a
form-factor decomposition which provides twelve independent
structure functions \cite{Pennington}. Only one of them needs
renormalization and corresponds to the term $\bar{\psi}A\!\!\!/\psi$
in the QED Lagrangian. We denote the projection of the QED vertex
function onto this form-factor as the amplitude $\in {\mathcal R}$
corresponding to this term in the Lagrangian.

A general amplitude $a\in {\mathcal A}$ is thus specified by a
particular form-factor and the quantum numbers of particles
participating in a scattering type experiment contributing to that
form-factor. It suffices to consider one-particle irreducible (1PI)
Green functions $G^{a}$ for each amplitude $a$. The knowledge of
these Green function suffices to determine all relevant physics by
standard techniques.

We assume that the amplitudes $a$ allow to specify an integer
$n=n(a)$ which gives the number of external legs. We let ${\mathcal
M}_a$ be the set of all 1PI graphs contributing to the amplitude
$a$. We remind the reader that a 1PI graph is a graph which remains
connected after removal of any one of its internal edges. By
$|\Gamma|$ we denote the number of independent loops in $\Gamma$. By
${\rm sym}(\Gamma)$ we denote the rank of the automorphism group of
the graph.

We then have in general \be G^{a}=1\pm\sum_{\Gamma \in {\mathcal
M}_{a}}\alpha^{|\Gamma|}\frac{\phi(\Gamma)}{{\rm
sym}(\Gamma)}=1\pm\sum_{k\geq 1}\alpha^k
\phi(c_k^a),\label{unrGF}\ee so that \be c_k^a=\sum_{\Gamma \in
{\mathcal M}_{a},\;|\Gamma|=k}\frac{\Gamma}{{\rm sym}(\Gamma)},
\label{series}\ee is the sum over all 1PI graphs of order $k$
contributing to the amplitude $a$. In the above, we take the plus
sign if $n(a)\geq 3$ and the minus sign for $n(a)=2$. We discard
tadpole amplitudes, $n(a)=1$, and vacuum amplitudes, $n(a)=0$.
 A main point of the subsequent
discussion will concern the structure of the sum \be
\Gamma^{a}=\One\pm\sum_{\Gamma \in {\mathcal
M}_{a}}\alpha^{|\Gamma|}\frac{\Gamma}{{\rm sym}(\Gamma)},\ee in the
above Green function $G^a$.

The set of amplitudes ${\mathcal A}$ decomposes for a renormalizable
theory into two disjoint subsets \be {\mathcal A}={\mathcal A}_+\cup
{\mathcal R}.\ee Here, to stress it once more, ${\mathcal R}$ is the
set of amplitudes for form-factors of one-particle irreducible
graphs which need renormalization. On the other hand, ${\mathcal
A}_+$ is the set of amplitudes which are overall finite.

Those amplitudes behave very differently. For an element $r \in
{\mathcal R}$ we can write \be \Gamma^r=\One\pm\sum_{\Gamma \in
{\mathcal M}^{r}}\alpha^{|\Gamma|}\frac{\Gamma}{{\rm
sym}(\Gamma)}=\One\pm\sum_{k\geq 1}\alpha^k B_+^{k;r}(\Gamma^r
Q^{n_rk}),\label{recursive}\ee where $B_+^{k;r}$ are Hochschild
one-cocycles as explained below. The quantity $Q$ is intimately
related to the notion of an invariant charge \cite{Gross}. For
momentum space Feynman rules $\phi$ we have \be \frac{\partial_L \ln
\phi(Q^{n_r})}{\partial L}{|_{L=0}}=\beta(\alpha),\label{qqq}\ee
where $\beta$ is the $\beta$-function of the theory which describes
the running of the charge $Q$. Hence $Q^{n_r}$ is a monomial in the
$\Gamma^r$, $r\in {\mathcal R}$, or their inverses. $n_r$ is an
integer such that $Q^{n_r}-\One$ is of order one in $\alpha$.
Finally, $L=\ln q^2/\mu^2$ sets the scale for that running charge.

From (\ref{recursive}), we have thus a self-similar recursive system
determining the formal sums $\Gamma^r$, $r\in {\mathcal R}$ in terms
of themselves and the action of suitable maps $B_+^{k;r}$. The study
of the maps $B_+^{k;r}$ is crucial for a QFT. The Hopf algebra
elements $B_+^{k;r}(\One)$ provide the skeleton graphs underlying
the DSEs and are the terms which drive the recursion which leads to
the full theory, and which connects QFT to motives \cite{BEK}.

In contrast, for $a\in {\mathcal A}_+$, we have \be
\Gamma^a=\One\pm\sum_{\Gamma \in {\mathcal
M}^{a}}\alpha^{|\Gamma|}\frac{\Gamma}{{\rm
sym}(\Gamma)}=\One\pm\sum_k B_+^{k;a}(M_a Q^{n_rk}),\ee where $M_a$
is another monomial in $\Gamma^s$, $s\in {\mathcal R}$, or their
inverses. Hence, amplitudes from the set ${\mathcal A}_+$ are
determined from the knowledge of the ones in ${\mathcal R}$.

Hence, we focus on amplitudes $r\in{\mathcal R}$  where it turns out
that the self-similarity and the properties of the Hochschild
one-cocycles $B_+^{k;r}$ are paramount to their understanding, as
well with respect to perturbative renormalization as with respect to
nonperturbative physics.
\subsection{The example of massless QED}
Massless quantum electrodynamics starts from the Lagrangian \be
{\mathcal L}_{\rm QED}=\bar{\psi}\left[\partial\!\!\!/+A\!\!\!/
\right]\psi+\frac{1}{4}F\cdot F.\ee

It has three monomials and as a renormalizable theory we hence can
work with a set \be {\mathcal R}_{\rm QED}=\{\v,\f,\g \},\ee in
standard notation.

The Green function for the vertex has an elaborate form-factor
expansion \cite{Pennington}, (with $n(\v)=3$), \be
G^{\v}_\mu=\phi(\Gamma^{\v})=\gamma_\mu
F_1(p_1^2,p_2^2,p_3^3,\alpha,\mu)+\cdots,\ee
 where it is only $F_1$ which needs renormalization. We let $P_1$ be
 the projector onto this form factor.

The unrenormalized Green functions for these monomials are then
given as \bea G^{\bar{\psi}A\!\!\!/\psi}_{\rm u} & = &
1+\sum_{\Gamma\in {\mathcal
M}_{ \v}}\alpha^{|\Gamma|}{\rm P_1}\phi(\Gamma),\\
G^{\bar{\psi}\partial\!\!/\psi}_{\rm u} & = & 1-\sum_{\Gamma\in
{\mathcal
M}_{ \f}}\alpha^{|\Gamma|}{\rm P_2}\phi(\Gamma),\\
G^{F\cdot F}_{\rm u} & = & 1-\sum_{\Gamma\in {\mathcal M_{
\g}}}\alpha^{|\Gamma|}{\rm P_3}\phi(\Gamma). \eea The projectors
$P_2$ onto the kinetic term of the inverse fermion propagator is
redundant in a massless theory and only needed in a massive theory,
and the projector $P_3$ onto the transversal part of the inverse
photon propagator is likewise redundant in the Landau gauge where
the tree level term remains transversal. We often write
$G^r=\phi(\Gamma^r)$ as a shorthand for such equations, with
suitable projectors understood in the application of the Feynman
rules $\phi$.

 Note that the Green function for the full vertex, containing eleven further amplitudes
 taken from the set ${\mathcal A}_+$, is obtained
by omitting the projector $P_1$ in the above. We stick to the notion
that each monomial in the Lagrangian has its own Green function,
while all other amplitudes belong to the set ${\mathcal A}_+$.

The goal is to calculate the corresponding 1PI Green functions order
by order in the fine-structure $\alpha$ of the theory, by applying
Feynman rules to these 1PI graphs of a renormalizable theory under
consideration. There are two problems here: each single graph is
mapped by the Feynman rules to an ill-defined quantity, and
furthermore, after labourosly eliminating these divergences, the
resulting series is not of the convergent type.

Progress with both problems is possible thanks to the algebraic
structures underlying Feynman graphs using the Lie and Hopf algebras
discussed below.

As we said before, 1PI Green functions are parameterized by the
quantum numbers, -masses, momenta, spin and such-, of the particles
participating in the scattering process under consideration.
Physicists denote propagating particles by lines, and the
perturbative expansion in terms of graphs is organized such that
external half-lines denote the particles parameterizing the
amplitude $a$ and the Green function $G^a$ under consideration,
while internal edges and vertices describe internal propagations and
vertices.

Note that the Lagrangian $L$ of massless quantum electrodynamics is
obtained accordingly as \be
L=\hat{\phi}(\f)^{-1}+\hat{\phi}(\v)+\hat{\phi}(\g)^{-1}=\bar{\psi}\partial\!\!\!/\psi
+\bar{\psi}A\!\!\!\!/\psi+\frac{1}{4}F\cdot F, \ee where
$\hat{\phi}$ are coordinate space Feynman rules. Here, inversion
like $ \hat{\phi}(\f)^{-1}$ takes account of the fact that monomials
quadratic in the fields refer to inverse propagators, while $\f$ and
$\g$ refer to the free propagators of QED.

This is not to say that there are no other Green function in quantum
electrodynamics. But we focus here on the Green functions which need
renormalization, and this, for a renormalizable field theory, gives
us a finite set of terms to be considered. It is indeed the
recursive self-similar nature of amplitudes from the set ${\mathcal
R}$ which drives the need for renormalization. That the
unrenormalized amplitudes suffer from short-distance divergences is
a mere accident of perturbation theory which disappears once the
non-perturbative fixpoint equations of motion, the DSEs, have been
taken into account.

The unrenormalized momentum space Feynman rules $\phi$ assign to a
graph a function of the external momenta $\{p_f\}$, and other
quantum numbers assiggned to external legs, ($\Gamma^{[0]}$ and
$\Gamma^{[1]}=\Gamma^{[1]}_{\rm int}\cup \Gamma^{[1]}_{\rm ext}$
being the set of vertices $v$ and internal and external edges $e$ of
$\Gamma$) of the form \be \phi(\Gamma)(\{p_f\})=\int \prod_{v\in
\Gamma^{[0]}}\phi(v)\delta^{(4)}\!\left(\sum_{f\; {\rm incident }\;
v}k_f\right) \prod_{e\in \Gamma^{[1]}_{\rm int}}{\rm
Prop}(k_e)\frac{d^4k_e}{4\pi^2}.\label{phi}\ee They are determined
from the knowledge of the Feynman rules for interaction vertices $v$
and the knowledge of free covariances ${\rm Prop}(k_e)$ for each
internal edge $e$. In the above form (\ref{phi}) a Dirac mass \be
\delta^{(4)}\left(\sum_{f\in \Gamma^{[1]}_{\rm ext}}p_f\right),\ee
will factor out of the expression for momentum conservation for
$|\Gamma^{[1]}_{\rm ext}|=n(a)$ external momenta $p_f$, for
$\Gamma\in {\mathcal M}_a$.

As a result, formally the unrenormalized Green function $G^a_{\rm
u}$ is obtained as \be G_{\rm
u}^{r}(\alpha;\{p_f\};z)=\phi(\Gamma^{r})
\left(\alpha;\{p_f\};z\right),\ee where we have introduced a
suitably chosen regulator $z$, needed in perturbation theory but not
non-perturbatively.

Note that in (\ref{phi}) the four-dimensional Dirac mass for each
internal vertex  guarantees momentum conservation at each such
vertex and restricts the number of four-dimensional integrations to
the number of independent cycles in the graph. These integrals
suffer from UV singularities which render the integration over the
momenta in internal cycles ill-defined. We remind the reader that
the problem persists in coordinate space, where one confronts the
continuation of products of distributions to regions of coinciding
support. We restrict ourselves here to a discussion of the situation
in momentum space and refer the reader to the literature for the
situation in coordinate space  \cite{BergbKrEG}.

The problem of perturbative renormalization is to make sense out of
this situation term by term: We have to determine invertible series
$Z^{r}(\alpha,z)$ for all $r\in {\mathcal R}$, hence in the
parameters of the Lagrangian \be L=\sum_{r\in {\mathcal R}}
\hat{\phi}^{\pm 1}(r),\ee such that the modified Lagrangian \be
\tilde{L}=\sum_{r\in {\mathcal R}}Z^{r}(\alpha,z)\,\hat{\phi}^{\pm
1}(r),\label{mod}\ee produces a perturbation series in graphs which
allows for the removal of the regulator $z$. Again, $\hat{\phi}$ are
the Feynman rules in coordinate space, with the understanding that
they evaluate an amplitude $r$ to the corresponding tree level term.

Let us first describe how this transition is achieved using the Lie-
and Hopf algebra structure of the perturbative expansion which we
summarize below:

\begin{itemize} \item Decide on the free fields and local
interactions of the theory, appropriately specifying quantum numbers
(spin, mass, flavor, color and such) of fields, restricting
interactions so as to obtain a renormalizable theory.
\item Determine the Feynman rules for free propagators from free
field theory. The Feynman rules for interaction vertices then follow
from locality: the ability to compensate by local counterterms
actually fixes the structure of interaction vertices modulo the
absolute values of masses and charges which parameterize the chosen
theory.
\item Consider the set of all 1PI graphs with edges
corresponding to those free-field propagators. Together with the
vertices  this allows to construct a pre-Lie algebra of graph
insertions. Anti-symmetrize this pre-Lie product to get a Lie
algebra ${\mathcal L}$ of graph insertions and define the Hopf
algebra ${\mathcal H}$ which is dual to the enveloping algebra
${\mathcal U}({\mathcal L})$ of this Lie algebra.
\item Realize
that the coproduct and antipode of this Hopf algebra give rise to
the forest formula which generates local counterterms upon
introducing a suitable Rota--Baxter map, a renormalization scheme in
physicists' parlance.
\item Use the Hochschild cohomology of this Hopf algebra to show
that you can absorb singularities in local counterterms, hence in
the form described in (\ref{mod}) above.
\end{itemize}
This gives a rather satisfying account of renormalization theory. As
an added bonus, if you work with an an intermediate complex
regulator $z$ these steps can be summarized as to construct the
Birkhoff decomposition of the unrenormalized Feynman rules, regarded
as an element in ${\rm Spec}({\mathcal G})$, the character group of
this Hopf algebra \cite{CK1,CK2,CK3}. This settles perturbative
renormalization.

The situation is even better when we look at the structure of the
full non-perturbative Green functions and concentrate on the
short-distance behaviour. Here, a regulator is not at all needed.
Indeed, in recent years, we have learned how to make progress with
non-perturbative physics \cite{anatomy,BrK,BergbKr,houches}, and
continue as follows:
\begin{itemize}
\item Show that the elements $c_k^r$ form a sub Hopf algebra.
\item Determine the Hochschild one-cocycles $B_+^{k;a}$ for this sub
Hopf algebra from the primitive elements of the Hopf algebra.
\item Choose a basis of amplitudes such that each Green function
which needs renormalization only depends on a single scale $L$.
\item Construct the DSEs as fixpoint equations with the elements
$\phi(B_+^{k;a}(\One))$ as kernels corresponding to the Dyson
skeleton expansion.
\item For the set ${\mathcal R}$, determine the recursion relations
which follow from the renormalization group upon rescaling $L$,
applying the consequences of the representations of one-parameter
groups of automorphisms of the above Hopf algebras \cite{CK3} in
this set-up.
\item Solve the DSE for the remaining unknown anomalous dimensions.
\item Aim to establish functional equations which connect an
expansion in $\alpha$ to an expansion in $1/\alpha$.
\end{itemize}
Very recently we gained insight how to carry such a program through
\cite{KrY}, and will exhibit below one example where it is carried
to the end.

What is encouraging are the structural features one can establish
for any renormalizable field theory which this program exhibits.

A first observation is that non-perturbatively, no regulator is
needed. The anomalous dimensions of propagators and vertices at zero
momentum transfer self-regulate the theory. Similar ideas, for the
conformal case of a vanishing $\beta$-function only, have been
employed in the past, and are reviewed in this volume \cite{Todor1}.

A second feature is that upon organizing the DSEs in terms of
Hochschild one-cocycles, the algebraic structures of the forest
formulas remain invariant upon addition of more one-cocycles. This
allows for a much more justified expansion, in terms of Hochschild
cohomology, than the usual truncations done in DSEs. It goes hand in
hand with a decomposition of field theory into periods of more and
more complicated motives. In particular, the decomposition into the
one-cocycles $B_+^{k;r}(\One)$ and the restriction to a subset of
these is a factorization of field theory reminiscent of a
decomposition of a zeta function into its Euler factors. This rather
far-fetched analogy \cite{Krbz} will be elaborated elsewhere.

From the viewpoint of DSEs, the Birkhoff decomposition becomes a
decomposition into a homogenous part and an inhomogenity determined
by the chosen renormalization condition. It thus survives the
transition to non-perturbative physics.

If the charge $\phi(Q^{n_r})$ gives rise to a vanishing $\beta$
function, the corresponding Hopf algebra governing the DSEs becomes
cocommutative, the dual Lie algebra is abelian and we can solve
non-perturbatively by a scaling solution and a simple Mellin
transform of the one-cocycles $B_+^{k;r}$ \cite{BergbKr,linear}.

In the general case, starting from this Mellin transform, one
realizes that the renormalization group allows to work with a very
simplified form of the coproduct: we can project on both sides to
terms in the span of linear generators. Hence we find two extremely
useful simplifications: the restriction to sub Hopf algebras which
are much easier to define than the full graph algebras, and the
linearization in the use of the coproduct. Together with the fact
that the primitives $B_+^{k;r}(\One)$ exhibit conformal invariance
at the renormalization point, which severely restricts the form of
their Mellin transform, this goes a long way, I set my hopes for the
future, in understanding better the phenomenon of universality,
finding common anomalous dimensions in very different physical
systems. But before we can exhibit these features in the promised
example, we have to start with the structure of Feynman graphs.

\section{Lie- and Hopf algebras of graphs}
To capture the structure of a renormalizable quantum field theory,
we organize it in terms of graphs. More formally, these graphs
$\Gamma$ will index generators $\delta_\Gamma$ of a Hopf algebra, so
that as an algebra it is the free commutative algebra on generators
indexed by the 1PI graphs of the theory. We will also consider
pre-Lie and Lie algebras, with their generators $Z_\Gamma$ indexed
by the very same graphs. We will often simply write $\Gamma$ for
$\delta_\Gamma$ or $Z_\Gamma$ when the context is clear.

All algebras are supposed to be over some field $\mathbb{K}$ of
characteristic zero, associative and unital, and similarly for
co-algebras. The unit (and by abuse of notation also the unit map)
will be denoted by $\One,$ the co-unit map by $\bar{e}.$ Algebra
homomorphisms are supposed to be unital. A bialgebra
$\left(A=\bigoplus_{i=0}^\infty A_{i},m,\One,\Delta,\bar{e}\right)$
is called graded connected if $A_{i}A_{j}\subset A_{i+j}$ and
$\Delta(A_i)\subset\bigoplus_{j+k=i}A_j\otimes A_k,$ and if
$\Delta(\One)=\One\otimes\One$ and $A_0=k\One,$
$\bar{e}(\One)=1\in\mathbb{K}$ and $\bar{e}=0$ on
$\bigoplus_{i=1}^\infty A_i.$ We call $\ker\bar{e}$ the augmentation
ideal of $A$ and denote by $P$ the projection $A\rightarrow
\ker\bar{e}$ onto the augmentation ideal, $P=id-\One\bar{e}.$

Note that the augmentation ideal contains the quantum world: all
graphs containing loops belong to the augmentation ideal. The
classical world is captured by $A_0=k\One$.

Furthermore, we use Sweedler's notation $\Delta(h)=\sum
h^\prime\otimes h^{\prime\prime}$ for the coproduct. We define
\be{\rm Aug}^{(k)}= \left(\underbrace{P\otimes\cdots \otimes
P}_{k\;{\rm times}}\right)\,\Delta^{k-1},\; A\to
\{\ker\bar{e}\}^{\otimes k},\ee as a map into the $k$-fold
tensorproduct of the augmentation ideal. We let ${ A}^{(k)}=\ker{\rm
Aug}^{(k+1)}/\ker{\rm Aug}^{(k)}$, $\forall k\geq 1$. All bialgebras
considered here are bigraded in the sense that \be
A=\bigoplus_{i=0}^\infty A_{i}=\bigoplus_{k=0}^\infty { A}^{(k)},\ee
where $A_{k}\subset \oplus_{j=1}^k { A}^{(j)}$ for all $k\geq 1$.
$A_{0}\simeq { A}^{(0)}\simeq \mathbb{K}$.

The first construction we have to study is the pre-Lie algebra
structure of 1PI graphs.
\subsection{The Pre-Lie Structure}
We are considering 1PI Feynman graphs. 1PI graphs are naturally
graded by their number of independent loops, the rank of their first
homology group $H_{[1]}(\Gamma,\mathbb{Z})$. We already wrote
$|\Gamma|$ for this degree of a graph $\Gamma$.

As we stressed several times, a crucial notion is the external leg
structure of a graph. It determines the relevant amplitude to which
the graph contributes. The relevant contribution to the counterterm
in the Lagrangian is then obtained by evaluating the Feynman rules
on the tree level graph which corresponds to that amplitude. It is
thus profitable to have a map which assigns to a given graph that
tree level graph. This map ${\bf res}(\Gamma)$ is easily described:
for any graph, we shrink all its internal edges to a point, so that
a single vertex, with the external edges attached as half edges,
remains. This is the desired tree level graph. For a graph with two
external edges, the result is simply the two-point vertex of a
form-factor of a single inverse edge, otherwise, if $n(r)>2$, we get
an interaction term in the Lagrangian.

When we evaluate a graph by the Feynman rules, we hence obtain a
result in the form \be \phi(\Gamma)=\phi({\bf res}(\Gamma))X+Y,\ee
where $X$ is a superficially divergent Green function and $Y$
contributes to amplitudes in ${\mathcal A}_+$.

We will see below how to construct a QFT from primitive graphs
$\Gamma$. Primitive graphs have no divergent subgraphs which need
renormalization. For such graphs $\Gamma$, the above decomposition
reads \be \phi(\Gamma)=\phi({\bf res}(\Gamma))\left[
\frac{r}{z}\right]+\;{\rm finite \;terms}+Y,\label{resid}\ee for
some regulator $z$. Here, $r$ is the numerical residue of the graph.
Thus, \be \lim_{z\to 0} z\phi(\Gamma)=r \phi({\bf res}(\Gamma)),\ee
and hence the name residue also for the map which shrinks all
internal edges. Note that $|{\bf res}(\Gamma)|=0$. We emphasize that
the set ${\mathcal M}_r$ contains only 1PI graphs $\Gamma$ such that
$|\Gamma|>0$. The residue ${\bf res}(\Gamma)$ is not an element of
this set, and hence is no generator in our Hopf algebra. The
elements of degree zero are strictly given by the scalars so that we
have a connected Hopf algebra, which is justified by the very fact
that each Green function is a mere structure function corresponding
to an amplitude in the Lagrangian.

Having specified free quantum fields and local interaction terms
between them, one immediately obtains the set of 1PI graphs, and can
consider for a given external leg structure $r$ the set of graphs
with that external leg structure. For a renormalizable theory, we
can define a superficial degree of divergence using the first Betti
number, \be \omega= \sum_{r\in \Gamma^{[1]}_{\rm int}\cup
\Gamma^{[0]}}\omega_{r}-4|H_{[1]}(\Gamma,\mathbb{Z})|, \ee for each
such external leg structure: $\omega(\Gamma)=\omega(\Gamma^\prime)$
if ${\bf res}(\Gamma)={\bf res}(\Gamma^\prime)$, all graphs with the
same external leg structure have the same superficial degree of
divergence -the hallmark of a renormalizable theory-, and only for a
finite number of distinct external leg structures $r\in {\mathcal
R}$ will this degree indeed signify a divergence.

For the pre-Lie structure we define a bilinear operation \be\Gamma_1
* \Gamma_2=\sum_\Gamma n(\Gamma_1,\Gamma_2;\Gamma)\Gamma,\label{prelie}\ee where
the sum is over all 1PI graphs $\Gamma$. Here,
$n(\Gamma_1,\Gamma_2;\Gamma)$ is a section coefficient which counts
the number of ways a subgraph $\Gamma_2$ can be reduced (by
shrinking it to its residue in the above sense) to a point in
$\Gamma$ such that $\Gamma_1$ is obtained. The above sum is
evidently finite as long as $\Gamma_1$ and $\Gamma_2$ are finite
graphs, and the graphs which contribute necessarily fulfill
$|\Gamma|=|\Gamma_1|+|\Gamma_2|$ and ${\bf res}(\Gamma)={\bf
res}(\Gamma_1)$, as ${\bf res}(\Gamma_1\star \Gamma_2)={\bf
res}(\Gamma_1)$ by construction.

 One then has:
\begin{theorem}
The operation $\ast$ is pre-Lie: \be [\Gamma_1\ast \Gamma_2]\ast
\Gamma_3  -  \Gamma_1\ast[\Gamma_2\ast \Gamma_3] = [\Gamma_1\ast
\Gamma_3]\ast \Gamma_2  - \Gamma_1\ast[\Gamma_3\ast \Gamma_2].\ee
\end{theorem}
Together with the corresponding Lie algebra ${\mathcal L}$ one is
led to consider the dual of its universal enveloping algebra
${\mathcal U}({\mathcal L})$ using the theorem of Milnor and Moore
and the above grading by the loop number. This graded dual, obtained
from the usual Kronecker pairing,  is a Hopf algebra ${\mathcal
H}(m,\One,\Delta,\bar{e})$ which is commutative but not
co-commutative. ${\mathcal
 H}$ is a graded commutative Hopf algebra which suffices to
describe perturbative renormalization theory \cite{K1,CK1,CK2}.

\subsection{Multiplicative subtraction}
The above algebra structures are available once one has decided on
the set of 1PI graphs of interest, delivering the renormalization of
any such chosen local quantum field theory. As to be expected, gauge
theories provide particular properties with respect to the
appearance of sub Hopf algebras which explain the Slavnov Taylor
identities for the couplings \cite{anatomy}, while the identities
related to kinematics of Green functions expressing transversality
of physical degrees of freedom for the gauge boson, are reflected in
the presence of Hopf ideals \cite{Sujl} and a semi-direct product
structure between radiation and matter \cite{dirktc}.

From the above, one-particle irreducible graphs $\Gamma$ provide the
linear generators $\delta_\Gamma$, with span ${\mathcal H}_{\rm
lin}={\rm span}(\delta_\Gamma)$, of the Hopf algebra ${\mathcal
H}=\oplus_{i=0}^\infty {\mathcal H}_i$. Disjoint union of graphs
provides the commutative product. We let $P$ be the projector into
the augmentation ideal ${\mathcal H}_{\rm aug}=\oplus_{i=1}^\infty
{\mathcal H}_i$ and $P_{\rm lin}$ be the projector into ${\mathcal
H}_{\rm lin}\subset {\mathcal H}_{\rm aug}$.

 Let now $\Gamma$ be a
1PI graph. We find the  Hopf algebra ${\mathcal H}$ as described
above to have a co-product explicitely given as $\Delta:{\mathcal
H}\to {\mathcal H}\otimes{\mathcal H}$:
\be\Delta(\Gamma)=\Gamma\otimes
\One+\One\otimes\Gamma+\sum_{\gamma{\subset}
\Gamma}\gamma\otimes\Gamma/\gamma,\ee where the sum is over all
unions of 1PI superficially divergent proper subgraphs (hence
subgraphs with residue in ${\mathcal R}$ each), and we extend this
definition to products of graphs so that we get a bi-algebra.

Having a co-product, two further structure maps of ${\mathcal H}$
are immediate, the co-unit and the antipode. The co-unit $\bar{e}$
vanishes on any non-trivial Hopf algebra element,
$\bar{e}(\One)=1,\,\bar{e}(X)=0$. The antipode is \be
S(\Gamma)=-\Gamma-\sum_{\gamma{\subset}
\Gamma}S(\gamma)\Gamma/\gamma.\ee

Note that for each term in the restricted sum $[P\otimes
P]\Delta(\Gamma)=\sum_i \Gamma_{(i)}^\prime\otimes
\Gamma_{(i)}^{\prime\prime}$ we have unique gluing data $G_i$ such
that \be
\Gamma=\Gamma_{(i)}^{\prime\prime}{\leftarrow_{G_i}}\Gamma_{(i)}^\prime,\;\forall
i.\label{glue}\ee These gluing date describe the necessary
bijections to glue the components $\Gamma_{(i)}^\prime$ back into
$\Gamma_{(i)}^{\prime\prime}$ so as to obtain $\Gamma$: using them,
we can reassemble the whole from its parts. Each possible gluing can
be interpreted as a composition in the insertion operad of Feynman
graphs \cite{dirkrev,MSS}. This gluing has a faithful representation
as an iteration of Feynman integrals, upon replacing the
$\delta$-function for momentum conservation at a given insertion
point by the expression for the inserted graph, and identifying the
continuous quantum numbers at external legs of the inserted graphs
with the momentum labels of the edges attached to the removed
vertex, in accordance with the gluing data. It is one of the
challenges with regard to analytic progress with Feynman graphs to
understand this operation as a generalization of the notion of
iterated integrals and their algebraic structure.

Disjoint scattering processes give rise to independent amplitudes,
so one is led to the study of characters of the Hopf algebra, maps
$\phi: {\mathcal H}\to V$ such that $\phi\circ m=m_V
(\phi\otimes\phi)$.

Such maps assign to any element in the Hopf algebra an element in a
suitable target space $V$. The study of tree-level amplitudes in
lowest order perturbation theory justifies to assign to each edge a
propagator and to each elementary scattering process a vertex which
define the Feynman rules $\phi({\rm \bf res}(\Gamma))$ and the
underlying Lagrangian, on the level of residues (in the graphical
sense above) of these very graphs. As graphs themselves are
constructed from edges and vertices, such residues indeed, one is
led to assign to each Feynman graph an evaluation in terms of an
integral over the continues quantum numbers assigned to edges or
vertices, which leads to the familiar integrals over momenta in
closed loops mentioned before, and hence leads to the Feynman rules
(\ref{phi}), $\phi\in {\rm Spec}({\mathcal G})$, as before.

Next, we choose a map  $R:V\to V$, from which we obviously demand
that is does not modify the UV-singular structure, and furthermore
that it obeys \be R(xy)+R(x)R(y)=R(R(x)y)+R(xR(y)),\label{RB}\ee
 an equation which guarantees the
multiplicativity of renormalization and is at the heart of the
Birkhoff decomposition which emerges below: it tells us that
elements in $V$ split into two parallel subalgebras given by the
image and kernel of $R$.  Such Rota--Baxter algebras play a role for
associative algebras which is similar to the role Yang--Baxter
algebras play for Lie algebras. The structure of these algebras
allows to connect renormalization theory to integrable systems
\cite{GF}. The situation is then remarkably similar to the
factorization and Birkhoff decomposition in many studies of
dynamical systems or problems in condensed matter theory, see for
example Korepin et.al.\ in this volume \cite{Korepin}. Also, most of
the results obtained initially for a specific renormalization scheme
like minimal subtraction can be obtained in general upon a
structural analysis of the corresponding Rota--Baxter algebras.

In renormalization theory we define a further character $S_R^\phi$
which deforms $\phi\circ S$ slightly and delivers the counterterm
for $\Gamma$ in the renormalization scheme $R$: \be
S_R^\phi(\Gamma)=-R m_V(S_R^\phi\otimes\phi\circ
P)\Delta=-R[\phi(\Gamma)]-R\left[\sum_{\gamma{\subset}
\Gamma}S_R^\phi(\gamma)\phi(\Gamma/\gamma)\right],\ee a slight
modification of the inverted Feynman rules \be \phi\circ
S=m_V(S\circ\phi\otimes\phi\circ P)\Delta=
-\phi(\Gamma)-\sum_{\gamma{\subset} \Gamma}\phi\circ S
(\gamma)\phi(\Gamma/\gamma).\ee Note that $S_ R^\phi\in {\rm
Spec}({\mathcal G})$ thanks to (\ref{RB}).

 The classical
results of renormalization theory follow using this group structure:
We obtain the renormalization of $\Gamma$ by the application of a
renormalized character \be S_R^\phi\star\phi(\Gamma)=m_V
(S_R^\phi\otimes\phi)\Delta,\ee and Bogoliubov's $\bar{R}$ operation
as \be\bar{R}(\Gamma)=m_V(S_R^\phi\otimes\phi)({\rm id}\otimes
P)\Delta(\Gamma)=\phi(\Gamma)+ \sum_{\gamma{\subset}\Gamma}
S_R^\phi(\gamma)\phi(\Gamma/\gamma),\ee so that we have \be
S_R^\phi\star\phi(\Gamma)=\bar{R}(\Gamma)+S_R^\phi(\Gamma).\ee Here,
$S_R^\phi\star\phi$ is an element in the group of characters
${\mathcal G}$ of the Hopf algebra, with the group law given by the
convolution
\be\phi_1\star\phi_2=m_V\circ(\phi_1\otimes\phi_2)\circ\Delta,\ee so
that the co-product, co-unit and co-inverse (the antipode) give the
product, unit and inverse of this group, as  befits a Hopf algebra.
This Lie group has the previous Lie algebra ${\mathcal L}$ of graph
insertions as its Lie algebra: ${\mathcal L}$ exponentiates to
${\mathcal G}$. This finishes perturbative renormalization theory.
Further results in particular with regard to the vicinity of
$\phi\in {\rm Spec}({\mathcal G})$ as tested by one-parameter groups
of automorphisms of the Hopf algebra were obtained in \cite{CK3}.
This illuminates in particular the renormalization group, whose
ubiquitous applications are discussed, for example, in this volume
\cite{McKeon}.

\section{The role of Hochschild cohomology and nonperturbative physics}
The Hochschild cohomology of the combinatorial Hopf algebras which
we discuss here plays three major roles in quantum field theory: it
allows to prove locality from the accompanying  filtration by the
augmentation degree coming from the kernels $\ker {\rm Aug}^{(k)}$,
it allows to write the quantum equations of motion in terms of the
Hopf algebra primitives, elements in ${\mathcal H}_{\rm lin}\cap
\{\ker {\rm Aug}^{(2)}/\ker {\rm Aug}^{(1)}\}$, and identifies the
relevant sub-Hopf algebras formed by time-ordered products. Before
we discuss these properties, let us first introduce the relevant
Hochschild cohomology \cite{CK1}.

\subsection{Hochschild cohomology of bialgebras} \label{Hochschild}

Let $(A,m,\One,\Delta,\epsilon)$ be a bialgebra, as before. We
regard  linear maps $L: A\rightarrow A^{\otimes n}$ as $n$-cochains
and define a coboundary map $b$, $b^2=0$, by
\begin{equation}
\label{Hscb} bL := ({\rm id}\otimes L)\circ
\Delta+\sum_{i=1}^n(-1)^i\Delta_i\circ L+(-1)^{n+1}L\otimes \One,
\end{equation}
where $\Delta_i$ denotes the coproduct applied to the $i$-th factor
in $A^{\otimes n}$, which defines the Hochschild cohomology of $A$.

For the case $n=1,$ (\ref{Hscb}) reduces to, for $L: A\rightarrow
A,$
\begin{equation}
\label{Hscb1} bL = ({\rm id}\otimes L)\circ\Delta-\Delta\circ
L+L\otimes\One.
\end{equation}
Note that for $h\in {A}^{(k)}$, we have $L(h)\in {A}^{k+1}$.

 The category of objects $(A,C)$ which consists of a
commutative bialgebra $A$ and a Hochschild one-cocycle $C$ on $A$,
and morphisms compatible with the cocycle actions,  has an initial
object $(\mathcal{H}_{\rm rt},B_+)$, where $\mathcal{H}_{\rm rt}$ is
the Hopf algebra of (non-planar) rooted trees and the closed but
non-exact one-cocycle $B_+$ grafts a product of rooted trees
together at a new root \cite{CK1}.

In Feynman graph Hopf algebras we will consider many one-cocycles
$B_+^{k;r}$. The closedness of these $B_+^{k;r}$ in the context of
Feynman graph Hopf algebras will turn out to be crucial for what
follows. Also, we note that such one-cocycles transpose to the
universal enveloping algebra on the dual side. With $B_+: {\mathcal
H}\to {\mathcal H}_{\rm lin}$ a one-cocycle, it turns out that by
the standard Kronecker pairing \cite{CK1,CK2}, the dual map
$B_+^\vee: {\mathcal L}\to U({\mathcal L})$ is a one-cocycle in Lie
algebra cohomology.

 \subsection{The roles of Hochschild cohomology}
The Hochschild cohomology of the Hopf algebras of 1PI graphs sheds
light on the structure of 1PI Green functions. To determine the
relevant Hochschild one-cocycles of a Feynman graph Hopf algebra
${\mathcal H}$, one determines first the primitives graphs $\gamma$
of the Hopf algebra, which by definition fulfill \be
\Delta(\gamma)=\gamma\otimes \One+\One \otimes \gamma.\ee Using the
pre-Lie product above, one then determines maps (possibly first on
suitably chosen sub Hopf algebras based on graph which have $\gamma$
as a cograph) \be B_+^\gamma:{\mathcal H}\to {\mathcal H}_{\rm
lin},\ee such that \be B_+^\gamma(h)=B_+^\gamma(h)\otimes
\One+\left({\rm id}\otimes B_+^\gamma\right)\Delta(h),\label{hcs}\ee
where $B_+^\gamma(h)=\sum_\Gamma n_+(\gamma,h,\Gamma)\Gamma$. The
new section coefficients $n_+(\gamma,h,\Gamma)$ are closely related
\cite{anatomy} to the section coefficients (\ref{prelie}) we had
before, but are normalized so that (\ref{hcs}) holds.

Using the definition of the Bogoliubov map $\bar{R}$ this
immediately shows that \be \bar{R}(B_+^\gamma(h))=\int {\rm
D}_\gamma {\leftarrow_{G_i}}
\left[S_R^\phi*\phi(h)\right],\label{loca}\ee which proves locality
of counterterms inductively upon recognizing that $B_+^\gamma$
increases the augmentation degree. By $D_\gamma$ we denote the
measure assigned by the Feynman rules to the primitive $\gamma$, \be
\phi(B_+^\gamma(\One))=:\int D_\gamma.\ee The insertion of the
functions for the subgraph is achieved using the relevant gluing
data of (\ref{glue}), and the start of the induction on primitive
graphs, which belong to ${\mathcal H}^{(1)}$, is immediate by
Weinberg's theorem \cite{Weinberg}. We hence obtain, by applying the
Rota--Baxter map $R$ to $\bar{R}$ in (\ref{loca}), the crucial
statement of renormalization theory: the counterterm in perturbation
theory is obtained by replacing all subgraphs by their renormalized
expressions.

If we want to understand now why a field theory can be renormalized
by local counterterms, we have to understand that it can be
organized such that at each order all its contributions are in the
image of a Hochschild closed one-cocycle.

It turns out that this can be achieved already at the combinatorial
level. The sum over all graphs $\Gamma^r$ contributing to a given
amplitude can indeed be obtained as the solution to a fixpoint
equation which puts these sums (for all $r\in{\mathcal R}$) into the
image of those Hochschild one-cocycles $B_+^{k;r}$.

Indeed, to recover the quantum equation of motions, the DSEs, from
Hochschild cohomology, one proves that \be
\Gamma^{r}=\One\pm\sum_{\gamma,\;{\bf
res}(\gamma)=r}\frac{\alpha^{|\gamma|}}{{\rm
sym}(\gamma)}B^\gamma_+(X_\gamma),\ee where \be X_\gamma=\prod_{e\in
\gamma^{[1]}_{\rm
int}}\prod_{v\in\gamma^{[0]}}{\Gamma^{v}}/{\Gamma^{e}} =\Gamma^r
Q^{n_r|\gamma|},\ee in a renormalizable theory. Upon application of
the Feynman rules the Hopf algebra elements $B_+^\gamma(\One)$ turn
to the integral kernels of the usual Dyson--Schwinger equations. We
obtain the required recursive form (\ref{recursive}) of the DSEs for
$r\in{\mathcal R}$. This allows for new non-perturbative approaches
as exhibited in a moment.

The one-cocycles introduced above allow to determine sub Hopf
algebras of the form \be\Delta(c^{r}_k)=\sum_{j=0}^k
P^r_{k,j}\otimes c_{k-j}^{r},\ee where the $c_j^{r}$ are defined in
Eq.(\ref{series}) and the $P^r_{k,j}$ are polynomials in these
variables given below. These sub Hopf algebras do not necessitate
the considerations of single Feynman graphs any longer, but allow to
establish renormalization directly for the sum of all graphs at a
given loop order.

They hence establish a Hopf algebra structure on time-ordered
products in momentum space. For theories with internal symmetries
one expects and indeed finds that the existence of these
sub-algebras establishes relations between graphs which are the
Slavnov--Taylor identities between the couplings in the Lagrangian
\cite{anatomy}.
\subsection{QED as an example}
Let us present the DSEs for massless QED. Thanks to the
Ward--Takahashi identity we only need to give them for two elements
in ${\mathcal R}_{\rm QED}$, the vertex $\v$ and the photon
propagator $\g$.

We give them first in graphical form using the results of
\cite{BDK}. \be \qeddse.\ee All internal edges are full propagators.
Note that in the second equation we took a double derivative with
respect to the external momentum at the inverse photon propagator.
The corresponding Green function is indicated on the lhs. That makes
it effectively a four-point function with two zero-momentum external
photons. Our Feynman rules, incorporating our definition of an
amplitude, are normalized to evaluate the tree-level term to unity,
and we get an expansion in terms of 1PI four-photon amplitudes,
organized in a skeleton expansion in this equation.

The combinatorial DSEs are thus simply \be \Gamma^{r}  =
\One\pm\sum_{\gamma\in {\mathcal H}^{(1)}\cap {\mathcal H}_{\rm
lin},\;{\bf res}(\gamma)=r}
\alpha^{|\gamma|}B_+^\gamma(\Gamma^{r}Q^{2|\gamma|})=\One\pm\sum_{k\geq
1}\alpha^kB_+^{k;r}(\Gamma^rQ^{2k}),\label{dseqed}\ee for
$r\in\{\v,\g\}$. The sum is over all 1PI primitive graphs with the
desired residue $r$. Also, \be
Q=\frac{\Gamma^{\v}}{\Gamma^{\f}\sqrt{\Gamma^{\g}}}.\label{Qqed}\ee
 Furthermore, \be B_+^{k;r}=\sum_{\gamma\in{\mathcal H}^{(1)}\cap {\mathcal H}_{\rm lin},{\bf
res}(\gamma)=r,|\gamma|=k}\frac{1}{{\rm sym}(\gamma)}B_+^\gamma, \ee
and $B_+^\gamma$ glues vertex functions and propagator functions
into edges of $\gamma$ in accordance with Hochschild cohomology.

The sub Hopf algebra is given as follows \cite{anatomy}. Let us look
at the set ${\mathcal R}_{\rm QED}$. Each of the series
$\Gamma^r=\Gamma^r(\alpha)$ is a formal series in $\alpha$, with
coefficients $c_k^r$. The sought-after polynomial $P^r_{k,j}$ in the
variables $c_m^s$ is simply the Taylor coefficient of order $j$ in
the Taylor expansion in $\alpha$ of the expression \be \Gamma^r
Q^{2(k-j)},\ee using (\ref{Qqed}). This is nothing but the argument
of the one-cocycle in (\ref{dseqed}) and this is true in general.
There is a wonderful and completely general phenomenon appearing
here \cite{anatomy,BergbKr}: if we restrict the sum over
one-cocycles in (\ref{dseqed}) at some loop order, we obtain the
same sub Hopf algebra. Those sub Hopf algebras are completely
universal and independent of the order of the skeleton expansion we
take into account. This is in my opinion a first clue as to why
phenomena like universality can appear in recursive dynamical
systems, to which quantum field theories evidently belong, at least
in their short-distance sector.

To see this result in action, let us a look at a three loop example
for the QED vertex. \be\begin{array}{lrcc} \Delta c_3^{\v}  =  &\One
& \otimes & c_3^{\v}\nonumber\\ & +[4
c_1^{\f}+5c_1^{\v}+2c_1^{\g}] & \otimes & c_2^{\v}\nonumber\\
 & +
\left[2c_2^{\f}+c_2^{\g}+3c_2^{v}+3c_1^{\v}c_1^{\v} \right. & & \nonumber\\
 & \left. + 3c_1^{\f}c_1^{\f} +  c_1^{\g}c_1^{\g}
+6c_1^{\f}c_1^{\v}\right. & & \nonumber\\
 & \left. +3c_1^{\g}c_1^{\v}+2c_1^{\g}c_1^{\f} \right] & \otimes &
c_1^{\v}\nonumber\\
& +c_3^{\v} & \otimes & \One. \end{array}\ee

 If we now impose
the Hopf ideal generated by the Ward identities \cite{Sujl} the
above simplifies to \be\begin{array}{lrcc} \Delta c_3^{\v} = & \One
&
\otimes & c_3^{\v}\nonumber\\ & + \left[c_1^{\v}+2c_1^{\g}\right] & \otimes & c_2^{\v}\nonumber\\
& + \left[c_2^{\v}+c_2^{\g}+c_1^{\g}c_1^{\g}  \right] & \otimes &
c_1^{\v}\nonumber\\
 & + c_3^{\v} & \otimes & \One,
\end{array}\ee and the Hopf algebra structure were co\-commuta\-tive were it not
for the non-va\-ni\-shing beta function, which ensures non-vanishing
$ c_k^{\g}$.

Actually, taken into account the classical results of Baker, Johnson
and Willey \cite{BJW}, the Hopf algebra would be trivial were the
$\beta$-function to vanish. A thorough discussion of such structures
in QED and their connection to the representation theory of the QED
Hopf algebra will be given elsewhere \cite{dirktc}.

Taking the Ward identity into account, the combinatorial DSEs are
even simpler, as then \be Q=\frac{1}{\sqrt{\Gamma^{\g}}}.\ee In
particular, for the photon we get \be \Gamma^{\g} =
\One-\sum_{\gamma\in {\mathcal H}^{(1)}\cap {\mathcal H}_{\rm
lin},\;{\bf res}(\gamma)={\g}}
\alpha^{|\gamma|}B_+^\gamma\left(\left[
\Gamma^{\g}\right]^{1-|\gamma|}\right).\label{dseqedp}\ee Note that
for the one-loop skeleton we have $|\gamma|=1$, the argument of the
one-cocycle is  $\One$ trivially, and hence this term becomes an
inhomogenous part in the DSE, and provides  the fermion determinant
for the QED Lagrangian.

\section{The structure of Green functions}
It is now time to summarize the structure of Green functions. We
closely follow \cite{anatomy,BrK,KrY,BergbKr}.
\subsection{Green functions for ${\mathcal R}$}
For a given superficially divergent amplitude $r\in {\mathcal R}$ we
let $\Gamma^r$ be the sum \be \Gamma^r=\One\pm\sum_{{\bf
res}(\Gamma)=r} \alpha^{|\Gamma|} \frac{\Gamma}{{\rm
sym}(\Gamma)},\label{dsesum}\ee over all 1PI graphs $\Gamma$
contributing to that amplitude, with $\alpha$ a loop-counting small
parameter.  Projection onto suitable form factors $\phi(r)$ allows
the sum to start with unity, so that by application of the Feynman
rules, $\phi(\Gamma^r)$ is the corresponding structure function and
the Lagrangian $L$ is given by \be L=\sum_{r\in{\mathcal
R}}\hat{\phi}^{\pm 1}(r),\ee where we take the plus sign for
$n(r)\geq 3$ and the minus sign for $n(r)=2$.

We write \be \Gamma^r = \One\pm
B_+^r(\Gamma^r,Q(\{\Gamma^i\})),\label{primsum}\ee where the
Hochschild one-cocycle \be B_+^r(\Gamma^r,Q)=\sum_{k\geq 1}\alpha^k
B_+^{k;r}(\Gamma^r Q^{n_rk}),\ee is a sum of one-cocycles
$B_+^{k,r}$ and $Q^{n_r}$ is a monomial in the $\Gamma^r$: \be
Q^{n_r}=\frac{1}{\Gamma^{{\bf res}(\gamma)}}\frac{\prod_{v\in
\gamma^{[0]}}\Gamma^v}{\prod_{e\in\gamma^{[1]}_{\rm
int}}\Gamma^e},\ee for any $\gamma\in {\mathcal H}^{(1)}$ and
$|\gamma|=1$. In general, different such $\gamma$ have different
internal edges and vertices, and hence we obtain different
$Q=Q(\gamma)$. For $Q$ to be well-defined, we divide by the
relations which equate them all. This is the origin of the
Slavnov--Taylor identities for the couplings \cite{anatomy}.

The $B_+^{k,r}$ themselves are obtained from the skeleton graphs
$\gamma$ of the theory: \be B_+^{k,r}=\sum_{\gamma\in {\mathcal
H}^{(1)}\cap{\mathcal H}_{\rm lin},{\bf res}(\gamma)=r}
\frac{1}{{\rm sym}(\gamma)}B_+^{\gamma},\ee where the sum is over
all Hopf algebra primitives $\gamma$ contributing to the amplitude
$r$ at $k$ loops. These maps are defined to be closed Hochschild
one-cocycles on the sub Hopf algebra generated by their
concatentations and products \cite{anatomy,BergbKr}.

Effectively, (\ref{primsum}) reduces the sum (\ref{dsesum}) over all
graphs to a sum over primitive ones, making use of the recursive
structure of this fixpoint equation which determines the sum of
graphs which contribute to  a chosen amplitude. The sums involved
typically reflect the universal law of \cite{Y} and will be
discussed in detail in upcoming work.

The $c_j^r$ are the linear generators of a sub-Hopf algebra:
\begin{theorem}
i) There exists a charge $Q$ as above, maps $B^{k;r}_+$ and
polynomials $P^r_{k,j}$ such that \bea \Gamma^r & = & \One\pm\sum_k
\alpha^k
B^{k;r}_+\left(\Gamma^r Q^{n_rk}\right),\\
\Delta B_+^{k;r} & = &  B_+^{k;r}\otimes \One + \left({\rm
id}\otimes B_+^{k;r}\right)\Delta,\\
\Delta c_k^r & = & \sum_{j=0}^k P_{k,j}^r\otimes c^r_{k-j},\eea
which make the free commutative algebra in generators  $\{c^r_k\}$
into a sub Hopf algebra
${\mathcal H}_{\rm c}(\Delta,m,S,\epsilon)$ of the Feynman graph Hopf algebra.\\
ii)  The polynomials $P^r_{k,j}$ are the Taylor coefficients of
$\Gamma^rQ^{n_r(k-j)}$ (the expansion of the argument of the $k$-th
one-cocycle) to order $j$ in $\alpha$.
\end{theorem}
We call the sub Hopf algebra ${\mathcal H}_{\rm c}$ the Hopf algebra
of time-ordered products, as the sum of all graphs with the same
residue delivers precisely that.
 Feynman rules are then defined in accordance with the Hochschild
cohomology as iterated integrals \be \phi(\Gamma)=\int {\rm
D}_\gamma {\leftarrow_{G_i}} \phi(h),\ee on subgraphs $h$, using
(\ref{glue}).

For a study of the short-distance behaviour of QFT we have to look
at the amplitudes in ${\mathcal R}$. For such amplitudes, we can
make use of the scaling properties of our Hopf algebras wrt
one-parameter groups of automorphisms, and combine the sub Hopf
algebra structure above with the general results on complex graded
commutative Hopf algebras of \cite{CK2,CK3}.

We have to make sure that any Green function which is superficially
divergent and appears in the integrand generated by
$B_+^{k;r}(\One)$ only depends on a single scale. To achieve this in
accordance with our approach is a non-trivial demand for amplitudes
with $n(r)>2$. The corresponding vertex functions
$\phi_R(\Gamma^r)$, $\phi_R:=S_R^\phi*\phi$,  depend a priori on
$n(r)-1$ linear independent external momenta $p_f$. Those external
momenta will be internal momenta inside the DSEs, where they are to
be integrated out thanks to the underlying iterated structure of
those equations. Thanks to the fact that short-distance
singularities are local we can isolate them into vertex functions at
zero momentum transfer, and decompose a divergent amplitude further
into an amplitude which depends on a single scale, and an amplitude
in ${\mathcal A}_+$. The choice of such a decomposition corresponds
to a choice of a basis of primitive elements in the Hopf algebra to
work with, and an easy recursion argument \cite{dirkrev} shows that
we can choose an appropriate basis of such primitives so that we can
set up the DSEs such that all divergent subgraphs depend only on a
single scale.

Hence there exists a basis of graphs and external structures in the
Hopf algebra such that \be \phi_R(Q)=\phi_R(Q)(\alpha,L),\ee where
$L=\ln q^2/\mu^2$ is the single scale which determines the running
of the invariant charge. In this chosen basis, short distance
singularities are captured by Green functions which are functions of
two dimensionless variables $\alpha,L$, with a remarkable duality
between these two variables first observed in \cite{BrK}.

In perturbation theory the Feynman rules now allow us to write the
renormalized Green functions  (compare to the unrenormalized ones in
(\ref{unrGF})), in the momentum scheme so that $G_R^r(\alpha,0)=1$,
as \be G_R^r(\alpha,L)=\phi_R(\Gamma^r)=S_R^\phi
*\phi(\Gamma^r)=1\pm\sum_k \alpha^k\phi_R(c^r_k)(L). \ee We can expand
in a different manner \be G_R^r(\alpha,L)=1\pm\sum_{k\geq 1}
\gamma_k^r(\alpha)L^k,\label{Lexp}\ee and the renormalization group
dictates relations between the $\gamma_k^r$. We work them out in a
moment and note the renormalization group equations first: \be
\frac{\partial G_R^r(\alpha,L)}{\partial
L}-\left[n_e(r)\gamma_1^e(\alpha)+\beta(\alpha)\frac{\partial}{\partial\alpha}\right]
G_R^r(\alpha,L)=0.\ee Here, $n_e$ is the number of external legs of
type $e$ in the residue $r$, $e\in {\mathcal R}$, $n(e)=2$ and
$\gamma_1^e$ is the corresponding anomalous dimension, obtained by
taking a derivative in (\ref{Lexp}), \be \gamma_1^e:=\frac{\partial
G^e(\alpha,L)}{\partial L}|_{L=0}.\ee $\beta(\alpha)$ is the
$\beta$-function corresponding to the charge $\phi_R(Q)$. It is
given through the anomalous dimensions in a simple form. If \be
Q^{n_r}=\frac{\prod_v [\Gamma^v]^{n_v}}{\prod_e
[\Gamma^e]^{n_e/2}},\ee then \be \frac{1}{n_r}\beta(\alpha)=\sum_v
n_v\gamma^v_1-\frac{1}{2}\sum_e n_e\gamma^e_1,\ee in accordance with
(\ref{qqq}). We let \be \frac{1}{n_r}{ \beta}_{\rm comb}:=\sum_v
n_v\Gamma^v-\frac{1}{2}\sum_e n_e\Gamma^e,
 \ee be the corresponding series in
$\alpha$ with coefficients in the Hopf algebra.

Next, we note that in the case of a linear DSE
\cite{BergbKr,linear}, we get \be
\partial_L \phi(Q)(L)=0.\ee As a result,  scaling  \be
G(\alpha,L)=e^{-\gamma(\alpha)L},\ee solves the linear DSE  so that
\be \gamma_k(\alpha)=\frac{\gamma_1(\alpha)^k}{k!}.\ee In such a
case, the corresponding Hopf algebra structure of the $c_k^r$ is
cocommutative and the dual Lie algebra structure abelian. The
violation of conformal invariance captured by a non-vanishing
$\beta$-function reflects itself in the non-vanishing commutators of
the Feynman graph Lie algebra.

A beautiful result of \cite{CK3} is the scattering formula, which
expresses higher poles in the regulator through iterated residues.
This translates into a statement on the coefficients of higher logs
in the leading log expansion of a quantum field theory. We are
assured that to get to these higher log contributions from a Hopf
algebra element, we simply have to apply the coproduct sufficiently
often so as to decompose it into the components in ${\mathcal
H}^{(1)} $. We then find the contribution to the $k$-th power of log
by the product of the residues of all those primitives.

Hence, to proceed in general we consider the map \be P^{(n)}_{\rm
lin}=\underbrace{P_{\rm lin}\otimes\cdots\otimes P_{\rm lin}}_{n\;
\rm times}\Delta^{n-1}\ee where $P_{\rm lin}$ is the projector into
the linear span of generators of the Hopf algebra. From
\cite{anatomy,BergbKr} we have: \begin{theorem} The linearized
coproduct is obtained as $$  P_{\rm lin}^{(2)}\Gamma^r = P_{\rm
lin}\Gamma^r\otimes P_{\rm lin}\Gamma^r+P_{\rm lin}Q^{n_r}\otimes
\alpha\partial_\alpha \Gamma^r,$$ where \be P_{\rm lin}Q^{n_r}={
\beta}_{\rm comb}.\ee\label{thm2}\end{theorem} This allows us to
understand the iterative structure of the next-to$\ldots$ leading
log expansion (\ref{Lexp}).

We define for $n>1$ \be
\sigma_n:=\frac{1}{n!}\;m^{n-1}\;\underbrace{\sigma_1\otimes\cdots
\otimes \sigma_1}_{n\;{\rm times}}\;\Delta^{n-1},\label{sig}\ee and
$\sigma_1$ is the residue defined by \be
\sigma_1=\partial_L\phi_R\left(S\star Y\right)(L)|_{L=0}.\ee

Actually, $\sigma_n$ evaluates to the coefficient of the $L^n$ term
in the evaluation of a Hopf algebra element by the renormalized
Feynman rules, by the scattering type formula \cite{CK3}.

We have \be h\not\in H_{\rm lin}\Rightarrow \sigma_1(h)=0,\ee so we
can use Theorem 2 and, by the above definition (\ref{Lexp}) of
$\gamma_k^r(\alpha)$, \be
\gamma_k^r(\alpha)=\sigma_k(\Gamma^r(\alpha)).\ee

Projection onto the linear generators delivers the desired formula
for the expansion in $L$, $\forall k\geq 2$: \be \gamma_k^r(\alpha)
= \frac{1}{k}\left[\gamma_1^r(\alpha)\gamma_{k-1}^r(\alpha) +\sum_j
s^j\gamma_1^j(\alpha)\alpha\partial_\alpha\gamma_{k-1}^r(\alpha)\right].\label{nextto}\ee

This gives us a second clue towards universality in field theory:
not only simplify the rather complicated graph Hopf algebras to the
rather simple Hopf algebras of time-ordered products, but from these
Hopf algebras we only need to remind ourselves of the simplest
linear part in them: the underlying complications of diffeomorphisms
of physical parameters only very mildly interfere with the
short-distance sector of a theory.

With the the above choice of basis we can now introduce the Mellin
transform by raising internal propagators ${\rm Prop}(k_e)$ with
momentum $k_e$ to non-integer powers $\rho$. A derivative wrt $\rho$
then amounts to the insertion of logarithmic corrections $\ln k_e^2$
into the Feynman integrals $\phi_R(B_+^{k;r}(\One))$, which is all
what is needed to proceed in view of the expansion (\ref{Lexp}) for
internal Green functions.

The DSEs turn into equations which determine $\gamma_1^r$ as we will
see in  a moment, while the further terms in the $L$ expansion are
determined from (\ref{nextto}) above. Green functions also have the
usual expansion in $\alpha$ which is triangular wrt $\gamma_k$: \be
\gamma_k^r(\alpha)=\sum_{j\geq k}\gamma_{k,j}^r\alpha^j.\ee We can
hence proceed to work out the recursion relations which express the
functions $\gamma_k^r$ through the functions $\gamma_1^r$ for $k>1$,
and turn the Dyson--Schwinger equations into an implicit equation
which allows to determine  the sole unknown coefficients
$\gamma_{1,j}^r$ from the knowledge of the above Mellin transforms.

Before we finally discuss an example, let us mention a third clue
towards universality: the fact that by construction the above
primitives are invariant under conformal inversion severely
restricts the form of the Mellin transform and analytic differences
can to a large extent be compensated by a redefinition of the
relevant couplings. We hope to have some more concrete results along
those lines in the future.

\subsection{A simple example} For concreteness, we consider massless
Yukawa theory and consider all self-iterations of the one-loop
massless fermion propagator, with subtractions in the momentum
scheme at $q^2=\mu^2$. Our Green function is an inverse propagator
with momentum $q$ and a function of two variables $a$ and $L=\ln
q^2/\mu^2$. We ignore radiative corrections at the bosonic line and
also at vertices, so the set ${\mathcal R}$ has a single element and
the superscript $r$ is suppressed henceforth. We restrict ourselves
to a single element in ${\mathcal H}^{(1)}\cap {\mathcal H}_{\rm
lin}$ and hence to a single one-cocycle $B_+$. We rederive the
results of \cite{BrK} for this case.

We write the perturbative series for the Dyson--Schwinger equation
as \be X(a)=\One-a B_+\left(\frac{1}{X(a)}\right),\ee where $
\phi(B_+(\One))$ provides the one-loop self-energy integral to be
iterated. Note that upon setting $X(a)=\One-\underline{X}(a)$, this
is the equation for the self-energy $\underline{X}(a)=-P_{\rm
lin}X(a)$ studied in \cite{BrK}.

We have \be Q=1/X^2\rightarrow P_{\rm lin}(Q)=-2\underline{X}(a),\ee
and find the linearized coproduct \be P^{(2)}_{\rm lin}X(a) = P_{\rm
lin}X(a)\otimes (P_{\rm lin}-2a\partial_a)X(a).\ee This is
Proposition 1 of \cite{BrK} and we also get
\begin{theorem}
The next-to next-to$\ldots$ leading log expansion in $L$ is given
through the anomalous dimension $\gamma_1(a)$ as \be
\gamma_k(a)=\frac{1}{k}\gamma_1(a)(1-2a\partial_a)\gamma_{k-1}(a).\ee
\end{theorem}
This is Proposition 2 of \cite{BrK}. As a result, can work out with
ease the recursions which express $\gamma_k$, $k>1$ through the
Taylor coefficients of $\gamma_1$.

Such recursions are obtained for any non-linear DSE by iterating
Theorem \ref{thm2}. We observe that we only need the cocommutative
part in the determination of the coproduct as is evident from the
definition (\ref{sig}) of $\sigma_k$, $k>1$. The non-cocommutative
part is always of lower degree in $L$ in the obvious filtration by
$L$.

 It remains to understand how to
compute $\gamma_1(\alpha)$. We proceed here by the Mellin transform.
It reads \be F(\rho)  =  \frac{1}{q^2}\int d^4 k \frac{k\cdot
q}{[k^2]^{1+\rho}(k+q)^2}{|_{q^2=1}} =\frac{1}{\rho(\rho-2)}  =:
\frac{r}{\rho}+\sum_{i\geq 0}f_{i}\rho^i. \label{Mellin}\ee Let us
introduce a short hand notation as \be \gamma\cdot
U:=\sum_{k=1}^\infty \gamma_k(\alpha) U^k.\ee The Dyson--Schwinger
equation becomes \be \gamma\cdot L=\alpha (1+\gamma\cdot
\partial_{-\rho})^{-1}[e^{-L\rho}-1]F(\rho)|_{\rho=0},\label{dsem2}\ee where we evaluate the
rhs at $\rho=0$ after taking derivatives. The functional dependence
of the non-linear DSE on $XQ=X^{-1}$ reflects itself on the rhs.

The only unknown quantities in this equation are the Taylor
coefficients $\gamma_{1,j}$ which are implicitly defined through the
Taylor coefficients of the Mellin transform (\ref{Mellin}) above.

Taking a derivative of (\ref{dsem2}) wrt $L$ and setting $L$ to zero
allows us to read them off: \bea \gamma_1
 & = & \alpha(1+\gamma\cdot \partial_{-\rho})^{-1} \rho F(\rho)|_{\rho=0}\\
 & = & \alpha r +\alpha
\left(\sum_{k\geq 1}[\gamma\cdot\partial_{-\rho}]^k\right)
\left[\sum_{k=1}^\infty \rho^k
f_{k-1}\right]|_{\rho=0},\label{DSErec}\eea so $\gamma_{1,1}=r$
universally. The resulting recursions determine the non-perturbative
solution given in \cite{BrK}.

As a final remark, we mention that a functional equation can be
assigned to the solution of the DSE, relating a weak to a strong
coupling expansion making use of the analytic structure behind the
recursion above. It is reminiscent of a functional equation for a
$\zeta$-function in two variables for the function field case
\cite{KrY}.

Furthermore, in the general case one needs multivariate Mellin
transforms, which is an immediate source for the appearance of
transcendentals in solutions to DSEs \cite{KrY}. The same holds
actually for the related study of higher one-cocycles \cite{BEK}. A
few short remarks finish our exploration of DSEs.
\section{From Primitives to Motives}
At long last we arrive in understanding the role of primitives
$\gamma\in {\mathcal H}^{(1)}$ in field theory. Their Mellin
transforms provide the basic constituents of a field theory, with
each Taylor coefficient of such a transform an interesting period in
its own right.

Knowledge of these Mellin transforms is at the time of writing
sparse and restricted to low degrees in $\alpha$, even if we are
only to look at the the residue, the first Taylor coefficient in
such a Mellin transform. Indeed, in accordance with (\ref{resid}) we
find \be \phi(B_+^\gamma)(\One)=\frac{r_\gamma}{\rho}+{\rm finite\;
terms},\ee and our first task is to find the residue $r_\gamma$ for
primitive graphs $\gamma$.

Still, such numbers provide much to explore for a mathematician, and
lead deep into the territory of algebraic geometry and motivic
theory. Data on such numbers have been amply provided in
collaboration with David Broadhurst \cite{Number,BK12} and John
Gracey \cite{john,BGK}, and hide in many computations in particle
physics \cite{LL}.

 While it is
nice that those Feynman graphs could be given a motivic
interpretation in \cite{BEK}, we are apparently (rather
depressingly, the evaluation of the wheel with $n$-spokes studied by
us in that paper is already a formidable task in this respect) a far
cry from a full motivic classification of Feynman graphs, though it
almost certainly exists. The results of \cite{BEK} where a first
step in this direction. Whilst incomplete, they give hope for the
future. One proceeds by assigning to a graph $\Gamma$ a graph
polynomial and finds the relevant pairings which are responsible for
the period assigned to the desired residue to be determined from the
interplay between the zeros of the graph polynomial (the graph
hypersurface) and the simplex of integration. The existence of
subgraphs containing closed loops (albeit convergent) ensures that
this interplay has non-trivial homology, and gives the resulting
periods motivic interpretation in the simple cases we could study.

The message to be kept is that upon organizing the short-distance
singular sector in a self-consistent manner, we find that we can
solve the DSEs upon understanding the number-theoretic and motivic
nature of the elements $\phi(B_+^{k;r}(\One))$.

\section*{Acknowledgment} Countless discussions with David
Broadhurst and Spencer Bloch on these topics, and encouragement from
my students Christoph Bergbauer, Kurusch Ebrahimi-Fard and Karen
Yeats, and some proofreading by Abhijnan Rej, are gratefully
acknowledged.


\begin{thebibliography}{9}
\bibitem{anatomy}
 D.\ Kreimer,  {\it Anatomy of a gauge theory},
 Annals of Physics (2006), online first, in press, hep-th/0509135.
\bibitem{instanton} A.\ I.\ Vainshtein, V.\ I.\ Zakharov, V.\ A.\ Novikov, M.\ A.\
Shifman,
    {\it The ABC of Instantons}, Sov.\ Phys.\ Ushpekhi {\bf 25}
(1982), 195-215.
\bibitem{GF} K.\ Ebrahimi-Fard, L.\ Guo, {\it Rota--Baxter algebras in renormalization of perturbative quantum field theory}, this volume;
hep-th/0604116.
\bibitem{Enc} D.\ Kreimer, {\it The Hopf algebra structure of renormalizable quantum field theory}, in Encyclopedia of Mathematical Physics,
eds.\ J.-P.\ Francoise, G.\ L.\ Naber, S.\ T.\ Tsou. Oxford:
Elsevier 2006 (ISBN 978-0-1251-2666-3).
\bibitem{BrK}
  D.~J.~Broadhurst and D.~Kreimer,
  {\it Exact solutions of Dyson-Schwinger equations for iterated one-loop
  integrals and propagator-coupling duality,}
  Nucl.\ Phys.\ B {\bf 600} (2001) 403, hep-th/0012146.
\bibitem{KrY} D.\ Kreimer, K.\ Yeats,
  {\it An etude in non-linear Dyson-Schwinger equations,}
  arXiv:hep-th/0605096, Nucl.\ Phys.\ {\bf B} Proc.\ Suppl., in
  press.
\bibitem{BergbKr}
C.\ Bergbauer,  D.\ Kreimer, {\it Hopf Algebras in Renormalization
Theory: Locality and Dyson–Schwinger Equations from Hochschild
Cohomology}, in {\it IRMA lectures in Mathematics and Theoretical
Physics}, Vol.\ {\bf 10}, Physics and Number Theory, European
Mathematical Society, Eds. V. Turaev, L. Nyssen;  hep-th/0506190.
\bibitem{houches} D.\ Kreimer, {\it Factorization in quantum field theory: An exercise in Hopf algebras and
  local singularities,}
  arXiv:hep-th/0306020, contributed to {\t Les Houches School of Physics: Frontiers in Number Theory, Physics and Geometry,}
   Les Houches, France, 9-21 Mar 2003, in press.
\bibitem{linear} D.\ Kreimer, {\it An \'etude in linear Dyson--Schwinger
equations,} preprint, IHES/P-06-23. MPI-Math, Bonn, Proceedings, to
appear.
\bibitem{BEK} S.\ Bloch, H.\ Esnault, D.\ Kreimer, {\it On Motives Associated to Graph
Polynomials,} Comm.\ Math.\ Phys.\ {\bf 267} (2006) 181-225;
  arXiv:math.ag/0510011.
\bibitem{Pennington}
 A.~Kizilersu, M.~Reenders and M.~R.~Pennington,
  {\it One loop QED vertex in any covariant gauge: Its complete analytic
  form,}
  Phys.\ Rev.\ D {\bf 52} (1995) 1242
  [arXiv:hep-ph/9503238].
\bibitem{Gross} D.\ J.\ Gross, {\it Application of the renormalization group to high energy physics,}
in {\it Les Houches 1975: Methods in field theory,} edited by R.
Balian and J. Zinn-Justin (North-Holland/World Scientific,
Amsterdam/Singapore, 1976/1981).
\bibitem{BergbKrEG} D.\ Kreimer, C.\ Bergbauer,
{\it The Hopf algebra of rooted trees in Epstein-Glaser
renormalization,}
  Annales Henri Poincar\'e {\bf 6} (2005) 343
  [arXiv:hep-th/0403207].
  \bibitem{K1} D.\ Kreimer,   {\it On the Hopf algebra structure of perturbative quantum field
theories,}
  Adv.\ Theor.\ Math.\ Phys.\  {\bf 2} (1998) 303
  [arXiv:q-alg/9707029].
\bibitem{CK1} A.\ Connes, D.\ Kreimer,
 {\it Hopf algebras, renormalization and noncommutative geometry,}
  Commun.\ Math.\ Phys.\  {\bf 199} (1998) 203
  [arXiv:hep-th/9808042].
\bibitem{CK2} A.\ Connes, D.\ Kreimer
{\it Renormalization in quantum field theory and the Riemann-Hilbert
problem.
  I: The Hopf algebra structure of graphs and the main theorem,}
  Commun.\ Math.\ Phys.\  {\bf 210} (2000) 249
  [arXiv:hep-th/9912092].
\bibitem{CK3}
  A.~Connes, D.~Kreimer,
  {\it Renormalization in quantum field theory and the Riemann-Hilbert  problem.
  II: The beta-function, diffeomorphisms and the renormalization
  group,}
  Commun.\ Math.\ Phys.\  {\bf 216} (2001) 215, hep-th/0003188.
\bibitem{Todor1} I.\ Todorov, {\it Constructing conformal field theory models}, this volume.
\bibitem{Krbz}
  D.~Kreimer,
  {\it What is the trouble with Dyson-Schwinger equations?},
  Nucl.\ Phys.\ Proc.\ Suppl.\  {\bf 135} (2004) 238, hep-th/0407016.
\bibitem{Sujl}   W.~van Suijlekom,
  {\it The Hopf algebra of Feynman graphs in QED,}
   Lett.\ Math.\ Phys.\ {\bf 77} (2006) 265; arXiv:hep-th/0602126.
\bibitem{dirktc} D.\ Kreimer, {\it On QED}, in preparation.
\bibitem{dirkrev} D.\ Kreimer,  {\it Structures in Feynman graphs: Hopf algebras and
symmetries,}
  Proc.\ Symp.\ Pure Math.\  {\bf 73} (2005) 43
  [arXiv:hep-th/0202110].
\bibitem{MSS} M.\ Markl, S.\ Shnider, J.\ Stasheff, {\it Operads in Algebra, Topology and
Physics}, Oxford Univ.\ Press 2002.
\bibitem{Korepin} A.\ R.\ Its, B.-Q.\ Jin, V.\ E.\ Korepin, {\it Entropy of XY spin chain and block Toeplitz determinants}, this volume.
\bibitem{McKeon} D.\ G.\ C.\ McKeon, {\it Using the renormalization group}, this volume.
\bibitem{BDK}   D.~J.~Broadhurst, R.~Delbourgo and D.~Kreimer,
  {\it Unknotting the polarized vacuum of quenched QED,}
  Phys.\ Lett.\ B {\bf 366} (1996) 421
  [arXiv:hep-ph/9509296].
\bibitem{BJW}   M.~Baker, K.~Johnson,
  {\it Quantum electrodynamics at small distances,}
  Phys.\ Rev.\  {\bf 183} (1969) 1292.
\bibitem{Y}
J.\ Bell, S.\ Burris, and K.\ Yeats, {\it Counting Rooted Trees: The
Universal Law
  $t(n)  C \rho^{-n}  n^{-3/2}$},
Elect.\ J.\ Comb.\ {\bf 13} (2006) \# R63, math.CO/0512432.
\bibitem{Weinberg} S.\ Weinberg,   {\it High-energy behavior in quantum field
theory,}
  Phys.\ Rev.\  {\bf 118} (1960) 838.
\bibitem{Number} D.\ Broadhurst, D.\ Kreimer,
 {\it Knots and numbers in $\Phi^4$ theory to 7 loops and beyond,}
  Int.\ J.\ Mod.\ Phys.\ C {\bf 6} (1995) 519
  [arXiv:hep-ph/9504352].
\bibitem{BK12} D.\ Broadhurst, D.\ Kreimer,
  {\it Association of multiple zeta values with positive knots via Feynman
  diagrams up to 9 loops,}
  Phys.\ Lett.\ B {\bf 393} (1997) 403
  [arXiv:hep-th/9609128].
\bibitem{john} J.\ Gracey, {\it Prac\-ticalities of re\-nor\-ma\-li\-zing quantum field theories}, this volume; hep-th\-/0605\-037.
\bibitem{BGK} D.\ Broadhurst, J.\ Gracey, D.\ Kreimer,
   {\it Beyond the triangle and uniqueness relations: Non-zeta counterterms at
  large N from positive knots,}
  Z.\ Phys.\ C {\bf 75} (1997) 559
  [arXiv:hep-th/9607174].
\bibitem{LL} S.\ Weinzierl, {\it The art of computing loop integrals}, this volume; hep-th/0604068.




\end{thebibliography}
\end{document}